%% file: MAIN.tex
\documentclass[aps, pra, a4paper, showpacs, twocolumn, english, 10pt]{revtex4-2}

\PassOptionsToPackage{dvipsnames}{xcolor}
\input{preamble}

\begin{document}

\title{Flexible Qubit Allocation of Network Resource States}

\author{
Francesco Mazza$^{1}$, 
Jorge Miguel-Ramiro$^{2}$, 
Jessica Illiano$^{1}$, 
Alexander Pirker$^{2}$, \\
Marcello Caleffi$^{1}$ and Angela Sara Cacciapuoti$^{1}$} 

\thanks{Corresponding author: Angela Sara Cacciapuoti\\ angelasara.cacciapuoti@unina.it}

\author{Wolfgang D\"ur$^{2}$}

\affiliation{$^{1}$University of Naples Federico II, Naples 80125, Italy \\$^{2}$Universit\"at Innsbruck, Institut f\"ur Theoretische Physik, Technikerstra{\ss}e 21a, Innsbruck 6020, Austria}

\begin{abstract}
    \textbf{Abstract.} The Quantum Internet is still in its infancy, yet identifying scalable and resilient quantum network resource states is an essential task for realizing it. 
    We explore the use of graph states with flexible, non-trivial qubit-to-node assignments. This flexibility enables adaptable engineering of the entanglement topology of an arbitrary quantum network. In particular, we focus on cluster states with arbitrary allocation as network resource states and as a promising candidate for a \textit{network core}-level entangled resource, due to its intrinsic flexible connectivity properties and resilience to particle losses.
    We introduce a modeling framework for overlaying entanglement topologies on physical networks and demonstrate how optimized and even random qubit assignment creates shortcuts and improves robustness and memory savings, while reducing the worst-case hop distance between remote network nodes, when compared to conventional approaches.
\end{abstract}

\maketitle

\section{Introduction}

Quantum entanglement promises to revolutionize the concept of network connectivity upon which we rely nowadays, posing major and compelling avenues for the realization of a future Quantum Internet \cite{CacCalTaf-20, CalCac-25}.
Although we are far from having a protocol suite for the Quantum Internet \cite{PirDur-19, IllCalMan-22, CacIllCal-23}, there are several use cases and protocols that exploit entanglement, and in particular multipartite entanglement, and show how this resource revolutionizes the concept of network connectivity \cite{PirDur-18, RamPirDur-21, CalAmoFer-22}.
As an example and with reference to the quantum teleportation protocol \cite{BenBraCre-93, CacCalVan-20}, entanglement can act as a \textit{half-duplex unicast channel} -- i.e., a communication channel connecting one single source to one single destination, and operated in one direction at a time (see the terminology clarification at the end of this section) -- between any pair of network nodes that share part of an entangled state. This new form of connectivity arises from the sharing of an entangled state rather than from the existence of specific physical links. Referred to as \textit{entanglement-based connectivity} \cite{IllCalMan-22}, it allows to redesign the topology of the network beyond its physical connectivity restrictions by introducing entangled links between physically unconnected remote network nodes \cite{MazCalCac-25}. These links serve as quantum connectivity shortcuts and enrich the physical topology with an \textit{entanglement topology}, namely, the topology induced by the entangled links, jointly specified by the structure of the shared resource state and by the assignment of its qubits to the network nodes. This topology is layered above the physical one, redefining the concept of neighborhood in quantum networks \cite{CheIllCac-25}, an effect that has been characterized and simulated at the network level in \cite{MazZhaChu-24}.

\begin{figure*}[t]
    \centering
    \includegraphics[width=0.95\textwidth]{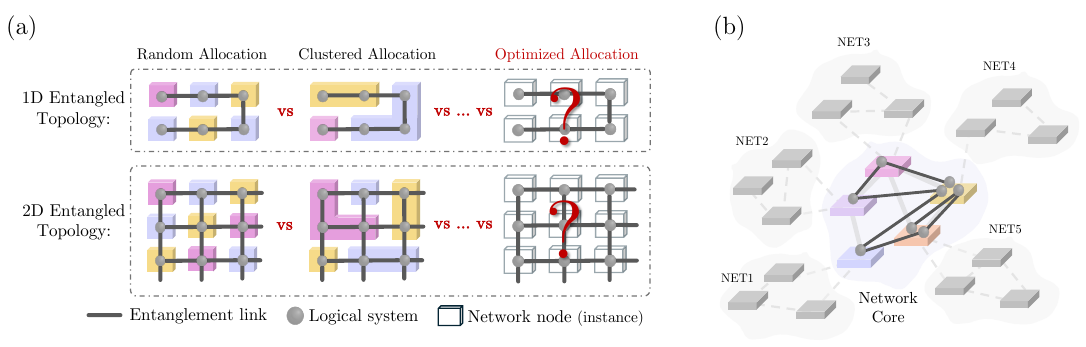}
    \caption{\justifying Schematic representation of our research problem. (a) Examples of qubit allocation strategies, illustrating the general allocation problem considered. Logical systems forming a cluster state are distributed across the network nodes. Each logical system, indicated with different colors, corresponds to an ensemble of physical qubits, resulting in a flexible and non-trivial allocation of entanglement resources, with the granularity of the logical system size. (b) Representation of large-scale quantum network comprising the interconnection of multiple small-scale quantum networks through the \textit{network core} \cite{CalCac-25}, the application that serves as a representative scenario to illustrate the benefits of flexible resource qubit allocation.}
    \label{fig:01}
    \hrulefill
\end{figure*}

In principle, an entangled state shared among network nodes allows the extraction of Bell states between remote pairs of nodes to meet \textit{communication requests}, i.e., demands for an entangled pair between two selected network nodes, to be served by consuming the shared resource.
The \textit{network resource}, namely the entangled state shared between the network nodes -- which, by oversimplifying, can be related to the number of Bell states (EPR pairs) concurrently obtainable from the same shared entangled state -- must be carefully designed to guarantee \textit{flexibility} and \textit{resilience} with respect to both traffic demands and network dynamism. The performance of such entanglement-based connectivity depends not only on which multipartite resource state is distributed among the nodes, but also on how the qubits of that state are assigned to the physical network nodes.
However, existing work on connectivity provisioning and distribution schemes specifies the resource state family together with a given qubit-to-node assignment, typically one qubit per node \cite{HahPapEis-19, FisTow-21}, or directly inherited from the resource state design \cite{FrePirDur-24, PirDur-21}, the specific resource exploitation protocol \cite{RuiWalDur-25}, or the physical layout \cite{MazCalCac-25}. Commonly, dedicated distribution protocols then treat the qubit-to-node assignment as an \textit{input} to the protocol, either fixed a priori or following a regular pattern, and optimize the generation and the delivery of the state so specified \cite{MeiMarGro-19, FanZhaGup-25, SenGooTow-25}.
Additionally, when transitioning from centralized to distributed scenarios, new layers of complexity emerge, and the design, allocation, and optimization of the network resource is highly non-trivial \cite{CicConPas-22,WanShiJi-25}. In this context, the central idea of this work is that the mapping of qubits of a shared resource state to network nodes, usually treated as a fixed implementation detail, can instead be optimized as a design variable (see Fig.~\ref{fig:01}a), enabling non-trivial entanglement topologies that improve both connectivity and resilience.
In realistic physical settings where multipartite states are usually generated locally and then distributed \cite{ThoRusMor-22, ThoRusMor-24}, the assignment of qubits to nodes can itself be regarded as a natural design variable.
By exploiting this, one can engineer an entanglement topology that departs from the physical layout and improves both connectivity and robustness to failures.

Among possible network resource candidates, relevant attention has been given to graph states \cite{HeiEisBri-04, HeiDurEis-06, VandeJHah-25}. More in detail, although the analysis of arbitrary Bell states extraction from graph states through LOCC is an NP-complete problem \cite{DahHelWeh-18}, some instances of graph states, as the linear (1D) and two-dimensional (2D) cluster states, stand up due to their regular structure and intuitive manipulation protocols \cite{HahPapEis-19, FrePirDur-24, deJJajTch-24}. These states have been shown useful to support on-demand fulfillment of network requests in both centralized and distributed settings \cite{MazCalCac-25, MazZhaChu-24, RuiWalDur-25}.

As mentioned above, flexibility and error resilience become mandatory features in distributed settings, as within a quantum \textit{network core} \cite{CalCac-25}, namely, the set of network devices and links acting as the backbone for interconnecting different networks, where no single node orchestrates the generation and manipulation of the resource. In fact, as exemplified in Fig.~\ref{fig:01}b, the nodes belonging to the network core exhibit the same role with no hierarchy and are responsible for the interconnection of multiple networks. Throughout this work, we use the network core as a primary use case to demonstrate the advantages of flexible qubit allocation within entangled resource states. While not the only possible application, the core exemplifies the challenges and benefits within distributed quantum networking.

The flexible allocation of resource qubits to network nodes entails a promising tool for building resilient and efficient quantum networks, see Fig.~\ref{fig:01}a. Rather than relying on a fixed or regular qubit-to-node mapping, non-trivial qubit assignment strategies allow to engineer the effective entanglement topology independently of the physical layout, enabling better adaptability to failures and communication demands. Such flexible qubit allocation is supported by realistic graph state generation tools, where multipartite entanglement is typically generated in some location \cite{Lu2007, Reiserer2015} and then distributed to the different nodes of the network \cite{ThoRusMor-22, ThoRusMor-24}.

In this paper, we explore this concept by focusing on a representative scenario: the \textit{quantum network core}. The network core refers to the set of nodes and links responsible for interconnecting multiple wide-area quantum networks covering different geographical areas, similar to their classical networks counterpart, as depicted in Fig.~\ref{fig:01}b. 
Given their primary role in interconnection of quantum networks, it is assumed that the core nodes are more powerful with respect to arbitrary quantum nodes.
Core nodes are expected to be equipped with sufficient quantum memory and connected by a small number of reliable and trusted links. This allows them to store multiple qubits and manage the preparation and distribution of the shared entangled resource.

We assume then that multiple qubits can be assigned to the same node, and joint access to and manipulation of all qubits within each single node is possible. This leads to a new, effective entanglement structure.

The core network resource can be generated and distributed through a \textit{proactive strategy} \cite{MurLiKim-16,RamPirDur-21, AbaCubMai-25}, where the entangled state is prepared in advance and refreshed periodically, independently of when the communication requests arrive and taking into account the current number of nodes of the network at generation time. As discussed in \cite{RuiWalDur-25}, this strategy allows immediate request fulfillment via local measurements and classical communication, avoiding delays due to on-demand resource preparation.

The \textit{research problem} we address can be stated as follows: model the communication scenario for a network comprising multiple interconnected nodes and design an entangled resource state and its qubit allocation strategy satisfying the following properties:
    \begin{itemize}
        \item [I)] Flexibility. The entangled resource state should dynamically accommodate node communication requests, i.e. the need of entangled pairs between selected end-nodes, accounting for topology changes;
        \item [II)] Resilience. The qubit allocation strategy must maintain operational and performance continuity in presence of node failures.
    \end{itemize}

The problem thus requires a framework that models the core communication scenario with a shared entangled resource, and allows arbitrary pairs of core nodes to extract Bell pairs by local manipulation of such resource state. In particular, we aim to define a \textit{resource state allocation strategy} that supports dynamic adaptation to physical topology updates and maintains network functionality under node failures or disconnections. This allocation becomes central in achieving both flexibility and resilience.

We highlight that our framework operates at the level of resource \textit{allocation}: it optimizes the assignment of the resource-state qubits to the network nodes, namely the allocation that the subsequent generation and distribution of the resource then realize. The generation and distribution themselves, addressed by dedicated entanglement-routing and distribution protocols \cite{PanKroHar-19, IneVarSca-23, AbaCubMai-25}, are deliberately abstracted, without altering the generality of our framework. 

To address this, we identify cluster states (see \textit{Preliminaries} in Results) as a suitable class of multipartite entangled states. In particular, we consider 1D and 2D cluster states as adaptable resources capable of fulfilling network requests on demand. Their structure enables the aggregation of multiple requests into a single flexible state (1D cluster states are well-suited candidates for multiple requests aggregation only if we consider that each network node is able to store multiple qubits; they also serve as flexible building blocks for more complex resource states), and their properties are well understood in terms of noise and losses \cite{HahPapEis-19, FrePirDur-24, FrePirDur-25}.
Indeed, when multiple qubits of the resource state are assigned to the same network node, joint operations on these qubits are possible locally. This leads to a new entanglement structure, enabling engineering and tailored optimization of the entanglement topology, adapting to communication requests.

In general, this class of states is capable of fulfilling requests by conforming to the physical network topology and establishing an entangled
path that extends beyond the originating quantum local area network (QLAN) \cite{MazCalCac-25, MazCalCac-24-QCNC}. 

The deployment of 1D and 2D cluster states in the core of a larger-scale quantum network promises to enable an efficient \textit{entanglement backbone} for agglomerating and routing multiple communication requests beyond localized quantum network clusters \cite{CalCac-25}.

We now clarify the terminology adopted throughout this work. In the context of classical communication networks, with \textit{unicast} communication is intended a one-to-one communication, namely, the transmission of information between one source and one and only one destination node. In contrast, a broadcast (multicast) communication is one-to-all (one-to-many), namely, the communication channel allows the transmission of the \textit{same} information, e.g., a given sequence of bits encoding the information, from one source to all nodes within the communication range. Furthermore, a full-duplex channel is a point-to-point communication system that connects two (or more) parties, allowing them to communicate in both directions simultaneously. Differently, a half-duplex system allows only one communication direction at a time \cite{Tan-10}. A simplistic example of a half-duplex channel is radio link between two parties, where both parties share the same frequency for transmission and reception, and within a given time interval, only one of them can transmit while the other receives. For a detailed discussion of the parallelism and differences between entanglement and classical communication systems we kindly refer the reader to \cite{IllCalMan-22}.

The main contributions of this work can be summarized as follows:
\begin{enumerate}
    \item[I)]  We consider non-trivial arbitrary qubit allocation for the network resource state and show the connectivity and robustness advantages of such an approach, and importantly, even with random allocation.
    We introduce a modeling framework and parameters for the use of graph states as network resources, i.e., shared quantum states, enabling flexible on-demand engineering of the network topology;

    \item[II)] We particularize the framework to the use of 1D and 2D cluster states as promising network resources to enable parallel and failure resilient communication. We analyze the features of the resulting network states in terms of overall number of hops (distance) required to communicate;

    \item[III)] We analyze the computational cost of the proposed optimization, by deriving the complexity of the allocation heuristic and by studying its convergence. This delimits the network scales at which the optimized allocation can be computed in time, and frames the role of the lightweight techniques -- such as the random allocation -- that can be employed during the optimization time;

    \item[IV)] We compare the optimized allocation of cluster state resource states with a random one, with a Bell-pair repeater chain, and with the straightforward \textit{all-to-all} communication scenario, whose maximum connectivity comes at a quadratic memory cost;

    \item[V)] We conduct an extensive resilience evaluation and show that an optimized allocation overall improves the resilience of the network to random nested node failures, with respect to other qubit allocation strategies. We also show that random allocation represents a solid and practical alternative when the computational time to determine the next assignment is limited.
\end{enumerate}

The remainder of this paper is organized as follows. In Results we introduce graph theory concepts and we provide definitions and tools for the manipulation of graph states in our system model. We then provide the mathematical framework and resource state design for the qubit allocation problem in the network core, and we delve into the performance analysis of the network core resource with respect to different key metrics such as the worst-case inter-node distance and number of vertex-disjoint paths. The optimal solution is then evaluated through extensive performance analysis with respect to node failure. In Methods we present the numerical methodology, including the complexity and the convergence analysis of the allocation heuristic. Finally, in Discussion we conclude the paper with a summary of the main results and insights on future works.

\section{Results}

\subsection{Preliminaries}

Before delving into the network architecture and the investigation on the entangled network resource, it is useful to introduce  definitions and concepts related to graph theory and graph states. These definitions are preliminary for the discussion of the framework of our model and the description of the problem of optimizing the resource state allocation.

Formally, a graph $G$ is represented as a pair consisting of two (finite) sets, $V$ and $E$, such that $G = (V,E)$. Here, $V$ is the set of vertices -- also referred to as \textit{nodes} -- with a cardinality of $|V| = n$. Meanwhile, $E$ defines the set of edges, which represents the connections among the vertices.

\begin{definition}[Path] 
	\label{def:01}
	An $\{u,v\}$-path is an ordered list $p_{\{u,v\}} = (v_1,v_2,\ldots,v_\ell)$ of distinct vertices in $V$ so that $u=v_1$, $v = v_\ell$ and $\{v_i, v_{i+1}\} \in E$ for any $i$. 
\end{definition}

\begin{definition}[Shortest path distance]
    Let $G$ be a graph and let $V(p)$ be the set of vertices visited by a path $p$ between two vertices $u,v \in V$. Given all the possible paths between $u$ and $v$, denoted with the set $P(u,v)$, the \textit{shortest-path} distance can be defined as follows:
\begin{equation}
    \label{eq:shortest_path_dist}
    d(u, v) = \min_{p \in P(u, v)} |V(p)| - 1.
\end{equation}
The quantity $d(u,v)$ is the number of hops required to reach node $v$ from node $u$.
\end{definition}

\begin{definition}[Vertex-Disjoint Paths]
    \label{def:vertex_disjoint}
    Let $G = (V, E)$ be a graph and let $u, v \in V$ be two distinct vertices. A collection of $k$ paths $\{p_1, p_2, \dots, p_k\}$ from $u$ to $v$ is said to be vertex-disjoint if:
    \begin{equation}
        V(p_i) \cap V(p_j) \subseteq \{u, v\}, \quad  \forall i \ne j.
    \end{equation}
    Within the considered set, the maximum number of such vertex-disjoint paths between $u$ and $v$ is denoted by $\kappa(u, v)$.
\end{definition}

A graph $G=(V,E)$ is \textit{connected} if, for each pair of vertices $u,v \in V$, there exists a $\{u,v\}$-path in $E$. 

\begin{definition}[Connected Component]
    \label{def:connected_component}
    Let $G = (V, E)$ be an undirected graph. A \emph{connected component} of $G$ is the maximal subset of vertices $K \subseteq V$ such that, for every pair $u,v \in K$, there exists a $\{u,v\}$-path in $E$, and there does not exist $v \in V \setminus K$ such that $K \cup \{v\}$  also satisfies this property.
\end{definition}

Note that the graph $G$ can be decomposed into a collection of connected components $K = \{K_1, K_2, \dots, K_r\}$ that forms a partition of the vertex set $V$. That is, $V = \bigcup_{i=1}^r K_i$, where each $K_i$ is a (disjoint) connected component, and $K_i \cap K_j = \emptyset$ for all $i \neq j$.

\begin{definition}[Colored Graph]
    \label{def:colored_graph}
    Let $G = (V, E)$ be a graph and let $\mathcal{C} = \{1, 2, \ldots, C\}$ be a finite set of colors. A \emph{colored graph} is a graph in which each vertex $v \in V$ is assigned a color $f(v) \in \mathcal{C}$ via a coloring function:
    \begin{align}
        f : V &\longrightarrow \mathcal{C}= \{1, 2, \ldots, C\}, \label{eq:coloring_func} \\
        f(v) &= c \in \mathcal{C}, \quad \forall v \in V. \nonumber
    \end{align}
    The function $f$  induces a partition of the vertex set into color classes: for each $c \in \mathcal{C}$, the set of vertices of color $c$ is defined as:
    \begin{equation}
        S_c = \{ v \in V \mid f(v) = c \} \subseteq V.
        \label{eq:colored_sets}
    \end{equation}
\end{definition}

Graph states represent an important subset of multipartite entangled states, widely investigated for their unique properties and their applications in quantum computation and communication scenarios \cite{RauBri-01, BarBirBom-23, HahPapEis-19, RuiRamWal-25}. A graph state, denoted by  $\ket{G}$, can be effectively described through well-known graph theory tools, given the straightforward correspondence between its entanglement interactions in the form of a graph. Formally:

\begin{definition}[Graph State]
\label{def:graph_state}
    A graph state $\ket{G}$ associated to the graph $G = (V,E)$ is defined as \cite{Gottesman1997, HeiDurEis-06}: 
\begin{equation}
    \label{eq:graph_state}
	\ket{G} = \prod_{\{u,v\} \in E} \texttt{CZ}_{uv}\ket{+}^{\otimes n},
\end{equation}
with $n = |V|$, $\ket{+} = \tfrac{1}{\sqrt{2}}(\ket{0} + \ket{1})$, and $\texttt{CZ}_{uv} = \text{diag}(1,1,1,-1)$ denoting the entangling controlled-Z gate applied to the qubits associated to the vertices $u$ and $v$, corresponding to an edge in the associated graph $G$.

Graph states are stabilizer states where $\ket{G}$ corresponds to the unique $+1$ eigenstate of the stabilizers:
\begin{equation}
    K_a = X_a \prod_{b \in N_a} Z_b,
\end{equation}
for all $a \in V$ and with $N_a$ denoting the set of neighbor qubits of $a$ in the associated graph $G$.
\end{definition}

A crucial property of graph states is the mapping between single-qubit Pauli measurements and associated graph operations.

\begin{definition}[Pauli Measurement on Graph States]
    \label{def:Pauli_measurements}
    Let $\ket{G}$ be an $n$-qubit graph state with associated graph $G = (V,E)$. The application of a Pauli measurement on the qubit corresponding to vertex $a \in V$ results, up to local unitaries, in a new graph state $\ket{G'}$ associated to the graph $G' = (V', E')$. Formally:
    \begin{equation}
        P_{\xi} ^{(a)}\ket{G} = \ket{\xi,\pm}^{(a)} \otimes U_{\xi,\pm}^{(a)}\ket{G'_{\xi}},
    \end{equation}
    where $P_{\xi} ^{(a)}$ is the Pauli operator acting on the qubit associated to the vertex $a$, $\ket{\xi,\pm}^{(a)}$ is the eigenstate on which the projection is applied and $U_{\xi,\pm}^{(a)}$ is the unitary operation corresponding to the measurement.
\end{definition}

The resulting graph state after the measurement $\ket{G'_{\xi}}$ -- up to a unitary operation $U_{\xi,\pm}$ -- is given by the application of local complementations $\tau(\cdot)$ and vertex deletions on the initial associated graph $G$, depending on the performed measurement \cite{HeiEisBri-04, HeiDurEis-06}:
\begin{align}
    \label{eq:Pauli_Z}
    \ket{G'_{z}} &= \ket{G - a},\\
    \label{eq:Pauli_Y}
    \ket{G'_{y}} &= \ket{\tau_a (G) - a},\\
    \label{eq:Pauli_X}
    \ket{G'_{x}} &= \ket{\tau_{b_0}(\tau_a(\tau_{b_0}(G)) -a)}, \, \text{with} \; b_0 \in N_a
\end{align}

Among graph states, cluster states emerge as a relevant instance, characterized by their cluster structures, that is, connected subsets of simple cubic lattices $\mathbb{Z}^d$ in $d\geq 1$ dimensions \cite{RauBroBri-03, BriRau-01, ZhoZenXu-23}.
The simplest cluster state is the linear, or 1D, cluster state. It can be expressed as a particular case of Eq.~\eqref{eq:graph_state} as
\begin{equation}
\label{eq:1Dcluster}
    \ket{L}_\text{N} = \prod_{i=1}^{\text{N}-1} \mathtt{CZ}_{i, i+1} \ket{+}^{\otimes \text{N}}.
\end{equation}

\begin{figure*}[t]
    \centering
    \includegraphics[width=0.85\textwidth]{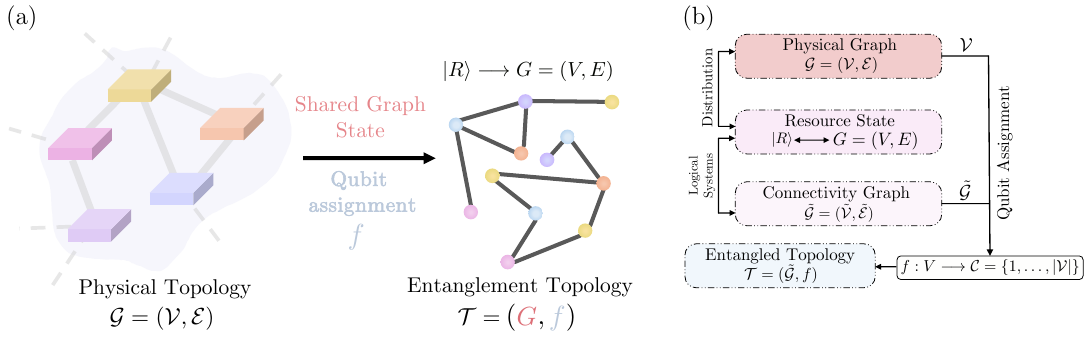}
    \caption{\justifying Schematic representation of our proposed framework for the description of resource state allocation and the entanglement topology of an arbitrary quantum network. (a) Example of physical topology $\mathcal{G} = (\mathcal{V}, \mathcal{E})$ and entanglement topology $\mathcal{T} = (G,f)$ after the share of a resource state $\ket{R}$, represented by a generic set of graph states. (b) Correspondence between physical graph, connectivity graph and entanglement topology of the network, through the application of a coloring function $f$ on the connectivity graph.}
    \label{fig:02}
    \hrulefill
\end{figure*}

Such simple 1D multipartite entangled state can be used as a promising building block for other graph states, thanks to its intuitive manipulation rules, and the possibility of generating such states in controlled environments \cite{HilLeoEis-23, ThoRusMor-22, ThoRusMor-24}. 
Notably, linear cluster states can also be built upon simple elementary merging operations at the network nodes, as described in the merging-based quantum repeater \cite{RuiRamWal-25}. 

The two-dimensional extension, i.e., the 2D cluster state, is characterized by a symmetric entangled structure corresponding to a two-dimensional rectangular lattice and has been widely investigated due to its universality as a resource for quantum computation, showcasing outstanding applications \cite{RauBri-01, RauBroBri-03, WalResRud-05, BriBroDur-09}. From a communication perspective, recent protocols have proposed the use of the symmetrical structure of these states to achieve multiple communication resources in parallel \cite{FrePirDur-24, FrePirDur-25}.
For instance, the \textit{zipper scheme} is a powerful tool for the multi-path generation of concurrent Bell states in a 2D cluster state \cite{FrePirDur-24}, and it is one of the most promising tools for the use of 2D cluster states as network resources. The scheme provides a scale-free network approach for generating a set of local Pauli measurements to perform on diagonal staircase-shaped paths within the lattice structure of the network resource.
This approach allows to generate such connections while preserving most of the remaining entangled structure of the cluster. It is worthwhile to anticipate that the generation of measurement patterns on the 2D entangled structure, as well as the possibility to use the zipper scheme, are at the basis of our proposed framework.

\subsection{Setting and general system model}

We introduce and explain the setting and system model considered, making use of graph states as shared entangled resources of the network and the tools introduced above. Furthermore, we discuss the key metrics and parameters employed in our framework.

The first step in outlining our communication scenario is to describe the network nodes and the underlying topology. While the physical topology corresponds to the actual geographical placement of the network components, the entanglement topology is defined by the distribution of an arbitrary graph state across the nodes. This entangled distribution effectively constructs an overlaying topology that can differ significantly from the physical one, circumventing spatial constraints, as illustrated in Fig.~\ref{fig:02}a.

Similarly to classical networks, the network physical topology can be represented by a graph $\mathcal{G}=(\mathcal{V},\mathcal{E})$, referred to as \textit{physical graph} where $\mathcal{V}$ is the set of the network nodes and $\mathcal{E}$ is the set of the physical channels between them. 

\begin{definition}[Number of nodes of the network]
    Given a graph $\mathcal{G}=(\mathcal{V},\mathcal{E})$ representing the physical topology of the network. The number of network nodes, also referred to as core nodes, is given by the cardinality of the vertex set: $C = |\mathcal{V}|$. Specifically, the set of nodes is defined as follows: $\mathcal{C} = \{c_1, \dots, c_{|\mathcal{V}|}\}$.
\end{definition}

In general, each node may store more than a single qubit, depending on the shared resource state and the available hardware.

\begin{definition}[Qubits per network node]
\label{def:qubits_per_node}
     The number of qubits stored at the $i$-th network node $c_i \in \mathcal{C}$ is denoted with $\square_{c_i}$.  
\end{definition}
Accordingly, the maximum number of qubits per network node in the network is given by:
\begin{equation}
    \square_c = \max_{c_i \in \mathcal{C}}{\square_{c_i}}.
\end{equation}

\begin{definition}[Resource state]
\label{def:resource_state}
    We define the resource state of the network as the shared multipartite entangled state $\ket{R}$ associated with the graph $G=(V,E)$.
\end{definition}

Network nodes can receive some network requests, such as the creation of Bell states, that has to be satisfied by exploiting the entangled network resource, which is set to be a shared graph state.

According to Def.~\ref{def:qubits_per_node}, a network node can store one or more entangled qubits belonging to the resource state. Hence, we can introduce a partition on the graph $G$ corresponding to the entangled resource, where each subset of vertices within the partition corresponds to a subset of entangled qubits belonging to the resource state. The graph obtained after the partition is a coarse-grained graph referred to as \textit{connectivity graph}, denoted with $\mathcal{\tilde G} = (\mathcal{\tilde V, \tilde E})$.

Here $\mathcal{\tilde V}$ denotes the set of logical vertices, which correspond in a one-to-many mapping to the vertices $V$ representing the entangled qubits of the resource state $\ket{R}$. Similarly, the edges of the connectivity graph $\mathcal{\tilde E}$ are in a one-to-many correspondence with the edges $E$, denoting entangled links of the resource state $\ket{R}$, as also represented in Fig.~\ref{fig:01}a.

\begin{remark}
    When allowing network nodes to store more than a single qubit, the graph associated to the resource state or the connectivity graph induced by the grouping of the physical qubits are not sufficient to describe the entanglement topology alone.
\end{remark}

The correspondence between the resource state, the physical graph and the connectivity graph is given by a coloring function $f$ (Def.~\ref{def:colored_graph}). More in detail, the coloring function expresses the assignment of the qubits of the shared entangled state to the physical network nodes $c \in \mathcal{C}$, i.e., the network nodes of the physical topology. As summarized in Fig.~\ref{fig:02}b, the model of the allocation of the resource state qubits to the nodes, allows us to define the \textit{entanglement topology} of the network:

\begin{definition}[Entanglement topology]
    \label{def:ent_topology}
    Given a graph $\mathcal{\tilde G}$ associated with a resource state and a coloring function $f$ acting on the graph $\mathcal{\tilde G}$, the entanglement topology is defined as the tuple $\mathcal{T} = (\mathcal{\tilde G}, f)$.
\end{definition}

\begin{remark}    
    The graph associated with the entanglement topology $\mathcal{\tilde G} = (\mathcal{\tilde V, \tilde E})$ is a \textit{connectivity graph} not necessarily identical to the graph $G=(V,E)$ associated with the graph state $\ket{G}$. 
    In other words, each vertex belonging to $\mathcal{\tilde V}$ corresponds to a \textit{logical system} of the shared entangled state. If each logical system is composed by a single qubit, then $\mathcal{\tilde G} \equiv G$.
\end{remark}

In general, as pictured in Figs.~\ref{fig:01}a and \ref{fig:02}b and better highlighted in \textit{Entanglement topology design}, logical systems correspond to an ensemble of physical qubits in the actual graph state $\ket{G}$ distributed among the network nodes and stored locally at the same network node. Similarly, different vertices of the entanglement topology can be associated with the same network node, thereby increasing the number of qubits allocated to the same network node, but keeping explicit the structure -- which, in the case of some resource states, such as cluster states, is geometric and recurrent -- of the entangled network topology. Note again that joint access to the qubits within the same node is assumed possible.

\subsection{Qubit assignment problem}

The allocation of an entangled resource state consists in the share of the entangled state according to a designated qubit assignment to the network nodes. Specifically, according to \textit{Preliminaries}, the assignment can be seen as a coloring function $f$ on the connectivity graph $\mathcal{\tilde G}$, thus defining the entanglement topology of the network $\mathcal{T} = (\mathcal{\tilde G},f)$. This function can also be seen as the output of an optimization problem, willing to improve the distances of the nodes in the entanglement topology as well as the communication capabilities of the shared resource.
In the following, we introduce the allocation strategies for 1D and 2D entanglement topologies.
First, we show an optimized qubit allocation strategy for our entanglement topology, capable of minimizing the worst-case shortest path between any pair of nodes.
Furthermore, we discuss how to design resilient resource states with desired 1D and 2D topologies, with optimized assignment.
Such optimization not only improves the efficiency of the resource's initial utilization but also enhances the intrinsic resilience of the network to random node failures.

A generalized entanglement topology can further benefit from flexible qubit allocation strategies for the shared resource state, where qubits can be flexibly distributed among the different parties to enhance overall resilience and connectivity features. This is described by the choice of the coloring function $f$, since different choices of coloring functions have a significant impact on the performance of the chosen topology, and thus it can be framed as an optimization problem. 
Our formulation for the optimized allocation strategy aims to minimize the distances between any pair of nodes in the entanglement network topology, which can be defined in terms of the number of hops between qubits (or logical systems, when referring to the graph $\mathcal{\tilde G}$) belonging to different network nodes.

The optimization problem belongs to the family of \textit{coloring problems}, where the network nodes can be seen as the colors to be assigned to each vertex of the connectivity graph $\mathcal{\tilde G}$.

Remarkably, the designed optimized allocation maximizes the fairness of the allocation of the entangled network resource, that is, each pair of randomly selected nodes that intend to communicate require a comparable number of hops in the lattice structure. In other words, the optimized allocation does not privilege any pair of nodes in the first utilization of the shared resource. Intuitively, shorter paths between network nodes imply less entanglement utilization in the shared resource (by assuming qubit decorations, discussed in \textit{Entanglement topology design}) and thus more useful entanglement for future requests.

The distance between two nodes in the entanglement topology impacts the length of the corresponding measurement pattern in the entangled lattice structure \cite{HahPapEis-19, RuiWalDur-25}, where higher distances correspond to lower quality of the extracted entanglement in presence of decoherence and gate imperfections.
A simple qualitative bound follows from the operational-fidelity model of \cite{MigIllMaz-25}, where the end-to-end fidelity can be estimated as $F(A,B) \lesssim F_0 e^{-t/\tau} q^{d(A,B)}$, with $q<1$ the per-operation quality, $\tau$ the memory coherence time, $t$ the storage time, and $F_0$ the initial quality of the stored resource. Consequently, the distance in the entanglement topology also bounds the worst-case achievable fidelity. However, the bound complements rather than replaces a full physical-layer performance evaluation.

Hence, the optimal solution of the assignment problem aims to minimize the worst inter-node shortest path. According to Eq.~\eqref{eq:shortest_path_dist},
by considering that the number of colors is set to the number of network nodes, for each node $c \in \mathcal{C}$, it is possible to define its \textit{inter-node distance} as follows:

\begin{definition}[Inter-Node distance]
\label{def:inter_color_distance}
Let $G=(V, E)$ be a colored graph with assignment function $f: V \rightarrow \mathcal{C} = \{1,\dots,C\}$ and let $S_c$ be the set of vertices assigned to node $c \in \mathcal{C}$.
The inter-node distance between two nodes $c, c' \in \mathcal{C}$, with $c' \neq c$, is defined as the minimum shortest path distance between any vertex in $S_c$ and any vertex in $S_{c'}$:
\begin{equation}
    \label{eq:inter_color_distance}
    d(c,c') = \min_{u \in S_c, v \in S_{c'}}  d(u,v).
\end{equation}
\end{definition}

Hence, the \textit{worst inter-node distance} -- to any other node -- is given by: 
\begin{equation}
    \mathcal{D}_c = \max_{c' \in \mathcal{C}: \, c' \neq c}  d(c,c').
\end{equation}

Furthermore, the objective function can be formulated as the minimization of the \textit{worst-case inter-node distance} over all nodes, thus lying in a classical \textit{min-max} problem:
\begin{equation}
    \text{obj:} \; \min_{f} \max_{c \in \mathcal{C}} \mathcal{D}_c.
\end{equation}

However, given the intrinsic complexity of finding an exact solution for arbitrary sized problems --  many variants of graph coloring, including optimization-based formulations, are known to be NP-hard \cite{Karp-2009, Gonzalez-1985} -- a heuristic solution can be employed.
Specifically, we use a simulated annealing algorithm for finding candidate solutions with $T_0 = 10$, a cooling rate of $0.99$ and $5000$ iterations. Simulated annealing represents here one of the possible implementations of the heuristic solution, adopted for its simplicity. Indeed, given the generality of the proposed formulation, any other search strategy can be equivalently employed, since the performance of the resulting entanglement topology and the related results depend on the quality of the returned allocation rather than on the procedure adopted to determine it. The results of the evaluation are discussed in \textit{Static performance evaluation} through a comparison with the random and clustered qubit allocation strategies represented in Fig.~\ref{fig:01}a.

A performance indicator used in the following for observing the effects of the proposed optimization is the number of vertex-disjoint paths (Def.~\ref{def:vertex_disjoint}) between different nodes. This is because each vertex disjoint path is related (as detailed below, depending on the presence of decorations and the internal structure of the logical systems) to the possibility of extracting dedicated Bell pairs and it is useful to estimate the parallel communication capabilities of the candidate assignment.  

Formally, vertex-disjoint inter-node paths are defined according to Def.~\ref{def:colored_graph} and by considering that each network node can be seen as a color in the colored graph:

\begin{definition}[Vertex-Disjoint Inter-Node Paths]
    \label{def:vertex_disj_inter_color_paths}
    Let $G=(V, E)$ be a colored graph with assignment function $f: V \rightarrow \mathcal{C} = \{1,\dots,C\}$ and let $S_c$ be the set of vertices assigned to node $c \in \mathcal{C}$.
    For two nodes $c, c' \in \mathcal{C}$, with $c \neq c'$, we define the support graph $G'=(V',E')$ such that two vertices are added to $G$ and connected to vertices assigned to $c$ and $c'$ respectively:
    \begin{align}
        V'&= V \cup \{s,t\}, \\
        E'&= E \cup \{(s,a), \, \forall a \in S_c\} \cup \{(t,b), \, \forall b \in S_{c'}\}.
    \end{align}
    We define the number of vertex-disjoint inter-node paths in $G$ according to the support graph $G'$:
    \begin{equation}
        \kappa(c, c') = \kappa(s,t),
    \end{equation}
    where $\kappa(s, t)$ is the maximum number of vertex-disjoint paths (Def.~\ref{def:vertex_disjoint}) between the additional vertices $s$ and $t$.
\end{definition}

\begin{remark}
    In particular, each path considered in $\kappa(c,c')$ corresponds to a path in $G$ from a unique vertex in $S_c$ to a unique vertex in $S_{c'}$, with no shared vertices among paths, including endpoints.
\end{remark}

We calculate the average number of vertex-disjoint inter-node paths, given any possible pair of nodes within the network.
Stemming from Def.~\ref{def:vertex_disj_inter_color_paths} and since each color $c \in \mathcal{C}$ corresponds to a network node, we can define the average number of vertex disjoint inter-node paths as:
\begin{equation}
        \bar \kappa = \frac{2}{|\mathcal{C}|(|\mathcal{C}| - 1)} \sum_{c, c' \, \in \, \mathcal{C}: \, c < c'} \kappa(c, c'),
\end{equation}
where we consider each possible unordered pair of nodes $c,c' \in \mathcal{C}$ with $c\neq c'$, since the paths are symmetric: $\kappa(c,c') = \kappa(c',c)$.

It should be noted that, given the connectivity graph $\mathcal{\tilde G}$, the value $\kappa(c,c')$ for two nodes $c$ and $c'$, is a proper indicator of the number of possible dedicated parallel connections available between two nodes. 
On one hand, each independent path corresponds to the possibility of obtaining a dedicated direct connection (a Bell state) by isolating the path with appropriate Pauli Z measurements to the neighbor qubits of the vertices of the paths. On the other hand, the independence of the paths in the connectivity graph guarantees that each path corresponds to a Bell state on the actual resource state, corresponding to a \textit{lower bound} on the number of parallel extractable connections between two arbitrary nodes. 
This is intuitively motivated by two factors: i) the use of the zipper scheme does not require the independence of the paths \cite{FrePirDur-24} and ii) the decorations of each vertex-disjoint path prevent the measurement of neighbor qubits belonging to adjacent independent paths, ensuring independent pattern isolation.

\subsection{Entanglement topology design}

The entanglement backbone design stems from the utilization of graph states, and specifically 1D and 2D cluster states as shared resource for the network core. 
As mentioned in the Introduction, the core nodes are assumed to be sufficiently powerful and reliable to handle the generation and proper distribution of the resource states. 
Specifically, the nodes belonging to the core network are interconnected through a simple and \textit{static} physical topology, represented in the form of a graph, where the physical links are well characterized and stable, and unlikely to change. In other words, given a narrow time range, the core network physical topology is static, meaning that it is unlikely that new nodes are added (the deployment of at least one new physical link is necessary) and unlikely to fail due to catastrophic errors. Clearly, whether a network node sharing an entangled state experiences a catastrophic failure or simply becomes unresponsive to a request, the entanglement involving that node can no longer be exploited. In this type of scenario, constant and thus proactive generation of multipartite states would ensure that requests are satisfied in parallel over time, adapting to current communication patterns. In the event of node failures, or most likely, unresponsive nodes, communication can still take place, thanks to the intrinsic flexibility and failure resilience properties of cluster states and careful allocation of qubits at network nodes.

In the following, we deepen the framework presented above, introducing its design parameters, by referring to cluster states as the resource state. As detailed in the following definition, cluster states can be generally represented by an associated graph $G$ with dimensions M$\times$N.

The associated graph $G=(V,E)$ and a coloring function $f$ fully describe a connectivity graph $\mathcal{\tilde G}$ corresponding to the entanglement topology of the network $\mathcal{T} = (\mathcal{\tilde G}, f)$, as defined in Def.~\ref{def:ent_topology}.

Accordingly, with $\mathcal{ \tilde G}$ we denote 1D or 2D lattice-shaped connectivity graphs, where the logical systems map to the physical resource $\ket{R}$ with associated graph $G$ and $\square_c$, i.e., the maximum number of physical qubits per network node, represents the network constraint for the construction of the entanglement topology. 
In order to describe the correspondence between the logical lattice graph $\mathcal{\tilde G}$ and the shared resource $\ket{R}$, the association between the logical and physical qubits must be formally described:

\begin{definition}[Parallelism factor]
    Given a graph $\mathcal{\tilde G}$ denoting the connectivity graph of the network, we define the (internal) parallelism factor $\mu$ as the number of physical qubits composing a logical system at the corresponding logical vertex $\tilde v \in \mathcal{\tilde V}.$
\end{definition}

\begin{remark}
    The parallelism factor reflects the internal properties of each logical system, indicating the minimal quantity of qubits to be allocated in batch, i.e., each time a network node is selected for the allocation. It also expresses the correspondence between the graph $G$ of the resource state $\ket{R}$ and the connectivity graph $\mathcal{\tilde G}$.
\end{remark}

\begin{figure}[t]
    \centering
    \includegraphics[page=1,width=0.42\textwidth]{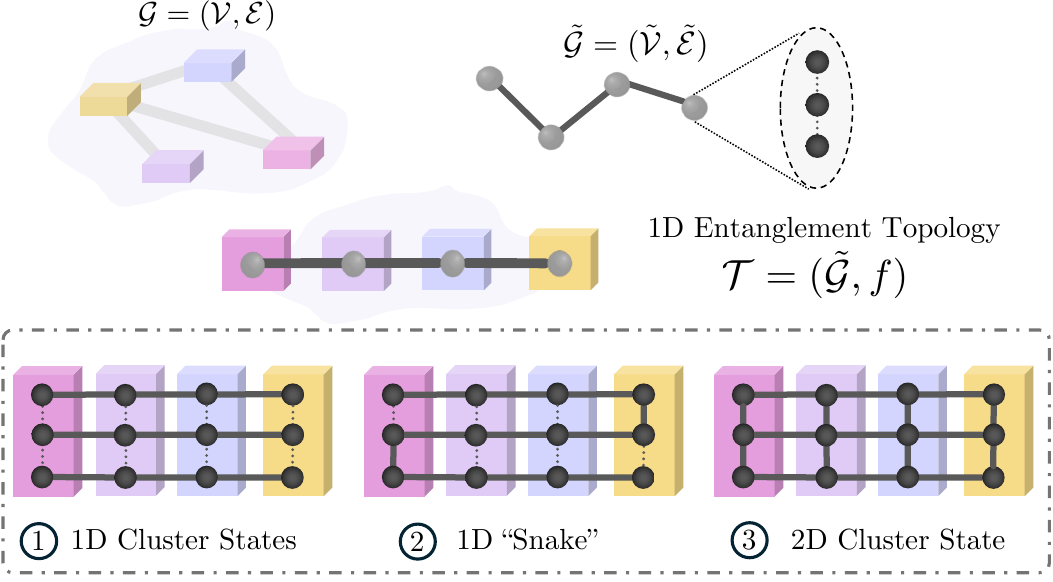}
    \caption{\justifying Example of 1D Entanglement topology with a given assignment $f$ and featuring a resource state $\ket{R} = \bigotimes_{m = 1}^{3} \ket{L}^{(m)}_\text{4}$ with $\text{N}=4$ and $\mu = 3$. The figure also shows some relevant internal structures of the logical systems, enabled by local operations at the nodes.}
    \label{fig:03}
    \hrulefill
\end{figure}

In the following, parallel stored qubits are represented in vertical disposition for the sake of clarity, as displayed in Fig.~\ref{fig:03}.
Moreover, since we focus on 1D and 2D entanglement topologies, the connectivity graph $\mathcal{\tilde G}$ can be characterized by its regular shape, reflecting the regularity of the underlying resource state;
we refer to $\mathcal{M} \times \mathcal{N}$ as the dimensions of the associated lattice-shaped connectivity graph $\mathcal{\tilde G}$, denoting the number of logical systems of the cluster states composing the generic M$\times$N resource state $\ket{R}$.

The simplest entanglement topology is given by the degenerate case of a $1\times\mathcal{N}$ lattice entangled topology, particularly relevant when it comes to near-term implementation of the considered communication scenario. Remarkably, a 1D entangled structure can be efficiently described with a flexible resource state exclusively composed of elementary 1D cluster states only and defining what we refer to as ``\textit{Snake}'' entanglement topology.
Accordingly, as represented in Fig.~\ref{fig:03}, the resource state corresponding to such an entanglement topology is given by ($\mu$ copies of) N-qubit 1D cluster states: 
\begin{equation}
    \label{eq:1Densemble}
    \ket{R} = \ket{L}^{(1)}_\text{N} \otimes \ldots \otimes \ket{L}^{(\mu)}_\text{N} = \bigotimes_{m = 1}^{\mu} \ket{L}^{(m)}_\text{N}, 
\end{equation}
and the assignment function $f$ is set to be the same for each of the $\mu$ cluster states.

Then, according to Def.~\ref{def:resource_state} and Eq.~\eqref{eq:1Dcluster}, a 1D entanglement topology can be defined by considering a N-qubit cluster state and described by the following set of parameters:
\begin{align}
\label{eq:snake_topo}
\mathcal{T}_{1,\mathcal{N}} = \mathcal{S_{N}} = \bigl(
    \text{N},
    \mu, 
    \square_c, 
    f \bigr),
\end{align}
where the coloring function $f$ defines how the resource state is allocated within the network nodes, and the dimensions $1\times \mathcal{N}$ denote the number of logical systems of the connectivity graph $\mathcal{\tilde G}$.

According to Eq.~\eqref{eq:colored_sets}, $f$ defines the sets of vertices belonging to the same node:
$S_c = \{ \tilde v \in \mathcal{\tilde V} \mid f(\tilde v) = c \}$,
where $S_c$ is the set of vertices of the entanglement topology assigned to node $c$.
Thus, the cardinality of the set $S_c$, for each node, is an indicator of the number of qubits stored locally at each network node:
\begin{equation}
    \square_{c_i} = \mu|S_c|, \quad \forall c \in \mathcal{C}.
\end{equation}

\begin{figure*}[t]
    \centering
    \includegraphics[width=0.85\textwidth]{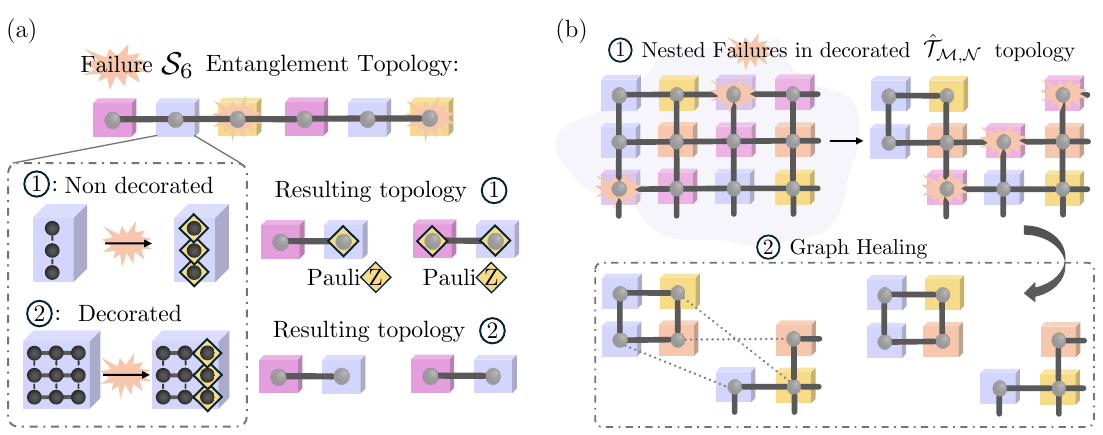}
    \caption{\justifying Node failures in a Snake entanglement topology $\mathcal{S_N}$ and -- decorated -- 2D entanglement topology $\mathcal{\hat T_{M,N}}$. (a) Example of node failure in a Snake topology $\mathcal{S_N}$, with core nodes $\{c_1, c_2, c_3\}$, parallelism $\mu = 3$, with and without qubit decorations; it highlights how qubit decorations allow for more remaining entangled links after a node failure. (b) Example of a sequence of node failures in a decorated $\mathcal{\hat T_{M,N}}$ entanglement topology with $5$ nodes; it highlights how (costless) local qubit operations can be used to restore connectivity between separate connectivity graph components after a sequence of node failures.}
    \label{fig:04}
    \hrulefill
\end{figure*}

As a result, as pictured in Fig.~\ref{fig:03}, every 1D cluster state in Eq.~\eqref{eq:1Densemble} follows the same qubit allocation scheme, resulting in a symmetric allocation, with exactly $\mu|S_c|$ qubits stored at each network node. Clearly, given this allocation rule of the qubits of the resource state, the connectivity graph $\mathcal{\tilde G}$ describing the entanglement topology is a simple linear graph with length $\mathcal{N}$. Here each logical system corresponds to $\mu$ parallel qubits, as also shown in Figs.~\ref{fig:03} and \ref{fig:04}a.

Notably, since the assignment fixes the same number of qubits per node for every strategy ($\square_{c_i}=\mu|S_c|$), the resource subsequently generated and distributed carries the same per-node load regardless of the chosen allocation. This means that the optimization selects \textit{which} qubits populate each node without changing \textit{how many} are delivered to it. Denoting by $\gamma_{c_i}$ the cost of delivering one qubit of the resource state to node $c_i$ -- set by the physical topology and by the distribution protocol, e.g., the number of required entanglement swapping operations -- the total distribution cost can be quantified as $\sum_i \square_{c_i} \gamma_{c_i}$, which is fixed by the per-node budgets and by the per-qubit delivery costs.
Provided that $\gamma_{c_i}$ is independent of \textit{which} logical qubit is delivered to node $c_i$, the allocation reshapes the entanglement topology at no additional distribution cost. This invariance follows from the adopted system model but, in general, routing and the distribution protocol may introduce a dependence on the identity of the delivered qubits.

Interestingly, as also represented in Fig.~\ref{fig:03}, if $\mu>1$, the shared resource can be equivalently seen as a 2D cluster state, serving as building block for the construction of 2D structures, where the qubits are arranged in a 2D grid with M rows and N columns. The 2D cluster state is obtained by the application of \textit{local operations} only at the network nodes, which are assumed to be \textit{costless} and not affected by any communication overhead.
This assumption reflects the powerful-core-node setting of the Introduction, while the impact of imperfect local operations and finite coherence budgets enters through the length of the measurement pattern required to serve a request, and is hence governed by the objective minimized in \textit{Qubit assignment problem}.

Thus, the entanglement topology is equivalent -- up to local operations only -- to a 2D cluster state with dimensions: M$\times$N $= \mu \times \text{N}$. Interestingly, the denomination \textit{Snake} comes from the fact that, according to the application of local operations only, it is possible to engineer the topology of the ensemble of linear cluster states described in Eq.~\eqref{eq:1Densemble} to a single, $\mu C$-qubit linear cluster state, with $\mu = 1$. 
Moreover, as better detailed in \textit{Resilience analysis}, 2D cluster states allow for better noise resilience in the case of node failures.

The definition of such an elementary resource state and the associated entanglement topology has different advantages.
The introduction of a flexible resource state, such as the one in Eq.~\eqref{eq:1Densemble} allows to rely on both the benefits of having multiple copies of shared linear cluster states and the possibility of locally engineering the entangled structure to an equivalent 2D cluster state.
Thanks to this flexibility, the entanglement topology can be engineered according to the communication needs, deciding to exploit the parallelized extraction enabled by the 2D cluster or an arbitrary simple merging based approach for single 1D cluster states.

\begin{remark}
    The simplest resource state defined in Eq.~\eqref{eq:1Densemble} is the building block resource state for every generic $\mathcal{T_{M,N}}$ entanglement topology. 
\end{remark}

Taking into account Def.~\ref{def:resource_state} and the designated resource state for our research problem (Eq.~\eqref{eq:1Densemble}), we formalize the generic 2D entanglement topology of our model as follows. 

\begin{definition}[2D entanglement topology]
We denote the entanglement topology as a 2D topology $\mathcal{T}_{\mathcal{M},\mathcal{N}}$, if the associated graph $\mathcal{\tilde G}$ is described by the following set of parameters:
\begin{align}
\label{eq:2D_topology}
\mathcal{{T}_{M,N}} = \bigl( 
    \text{M},
    \text{N},
    \mu,
    \square_c, f \bigr),
\end{align}
where $f$ denotes the assignment function, $\mu$ denotes the parallelism factor and $\mathcal{M}\times \mathcal{N}$ denote the dimensions of the connectivity graph $\mathcal{\tilde G}$.

\end{definition}

As in classical communication networks, quantum network nodes are also susceptible to failures, which may make them unable to process or fulfill entanglement requests from other nodes. Such failures can arise from a variety of causes, including catastrophic hardware failures or excessive requests. Regardless of the underlying cause, a well-designed multipartite entangled resource should exhibit a high degree of resilience to node disconnections. Specifically, the resource should maintain its ability to fulfill ongoing requests among the remaining operational nodes, ensuring that the maximum possible amount of useful entanglement is preserved and effectively distributed.

\renewcommand{\arraystretch}{1.2}

\begin{table}[t]
\centering
\caption{\justifying {Key parameters for the description of the network model.}}
\label{tab:key_parameters}

\fontsize{8pt}{8pt}\selectfont
\begin{tabular}{|c|p{0.6\linewidth}|}
\hline
\hline
\multicolumn{1}{|c|}{\textbf{Symbol}} & \multicolumn{1}{c|}{\textbf{Description}}   \\ 
\hline

$C$ & Number of core nodes of the network. \\ 

$\mathcal{C}$ & Set of core nodes of the network, also seen as the set of colors. \\

${\square_c}$ & Number of local qubits at each core node. \\

$\mathcal{G} = (\mathcal{V},\mathcal{E})$ & Graph representing the physical topology of the network. \\ 

$\mathcal{\tilde G} = (\mathcal{\tilde V},\mathcal{\tilde E})$ & Connectivity graph representing the entanglement topology of the network.\\

${\tilde v} \in  \mathcal{\tilde V}$ & Vertex of $\mathcal{\tilde G}$. Corresponds to a logical system of the shared resource state in the entanglement topology.\\

$\ket{R}$ & Resource state of the network. \\  

$f$ & Coloring function that describes the allocation of the qubits of the resource states to network nodes. \\

$\mathcal{T} = (\mathcal{\tilde G}, f)$ & Generic entangled network topology associated to the connectivity graph $\mathcal{\tilde G}$. \\

$S_c = \{ \tilde v \in \mathcal{\tilde V} \mid f(\tilde v) = c \}$ & Set of vertices of the entanglement topology with color $c$. \\

$\mu$ & Number of physical qubits corresponding to each logical vertex $\tilde v \in \mathcal{\tilde V}$. \\ 

M$\times$N & Dimensions 2D cluster state network resource. \\

$\mathcal{M}\times\mathcal{N}$ & Dimensions of the lattice-shaped entanglement topology $\mathcal{T}$. \\

$\mathcal{T_{M,N}}$ & 2D entanglement topology.\\

$\mathcal{\hat T_{M,N}}$ & Decorated 2D entanglement topology.\\

$\mathcal{S_{N}} \equiv \mathcal{T}_{1,\mathcal{N}}$ & Snake entanglement topology.\\

$\mathcal{\hat S_{N}}$ & Decorated Snake entanglement topology.\\

\hline
\hline
\end{tabular}
\end{table}

\begin{remark}
    The use of a cluster state as network resource intrinsically ensures adequate resilience to network failures. This is because, in contrast with other non-persistent quantum states (such as the GHZ state \cite{IllCacMan-21}), the loss of one of the parties does not affect the whole state and can be recovered by measuring out the neighbors of the lost particle. Thanks to the application of the measurement rules detailed in Def.~\ref{def:Pauli_measurements}, the qubits adjacent to a lost particle can be measured in the Z basis and removed from the cluster state by preserving the rest of the entangled structure, introducing a \textit{hole}. This effect can be enforced through qubit decoration at the nodes, which also minimizes the effective hole introduced in the cluster.
\end{remark}

If a node fails in our communication scenario, regardless of the cause, every qubit stored in that network node is lost in the entangled structure of the resource state. In other words, each of the qubits of the faulty node is not accessible and its neighboring qubits are measured in the Z basis (Def.~\ref{def:Pauli_measurements}, Eq.~\eqref{eq:Pauli_Z}), thus creating a \textit{hole} in the lattice entangled structure, as represented in Figs.~\ref{fig:04}a,~\ref{fig:04}b.

As depicted in Fig.~\ref{fig:04}a, in the specific case of a $\mathcal{S_{N}}$ topology, where the connectivity graph is linear, a hole splits the entanglement topology into separated graph components (Def.~\ref{def:connected_component}).
Hence, each failure interrupts any possible connection between different components. 

Crucially, local operations between qubits stored at the same network nodes are assumed to be costless.
As a consequence, if different logical vertices $\tilde v \in \mathcal{\tilde V}$ are allocated to the same network node then the holes created by network failures can be \textit{re-healed} by exploiting local operations between qubits stored locally at the same node, as schematically represented in Fig.~\ref{fig:04}b. As a result, the number of connected components of the connectivity graph can be reduced, and the connectivity restored.

The most relevant catastrophic consequence of network failures in a cluster state shared resource is given by a sequence of multiple node failures, represented in Fig.~\ref{fig:04}b. The need to trace out the neighbor qubits of each failed one causes a cascade of disconnections within the network topology, thus provoking the collapse of the entire entangled structure after a few nested failures.
Fortunately, it is possible to limit the number of disconnections caused by a sequence of network failures by considering the presence of additional support qubits stored at each network node. Such qubits are \textit{decorated vertices} \cite{MeiMarGro-19}, since they have no impact on the geometry of the network topology but only protect the original qubits from undesired cascade tracing outs, as depicted in Fig.~\ref{fig:04}a. 

\begin{remark}[Qubit decorations]
    The elementary resource state of Eq.~\eqref{eq:1Densemble} decorated with additional qubits, allows to reinforce the error resilience of the network -- for every general $\mathcal{T_{M,N}}$ entanglement topology -- at the cost of increasing the number of qubits per node $\hat \square_{c_i} \geq \square_{c_i}$.
\end{remark}

\begin{figure*}[t]
    \centering
    \includegraphics[page=1,width=0.8\textwidth]{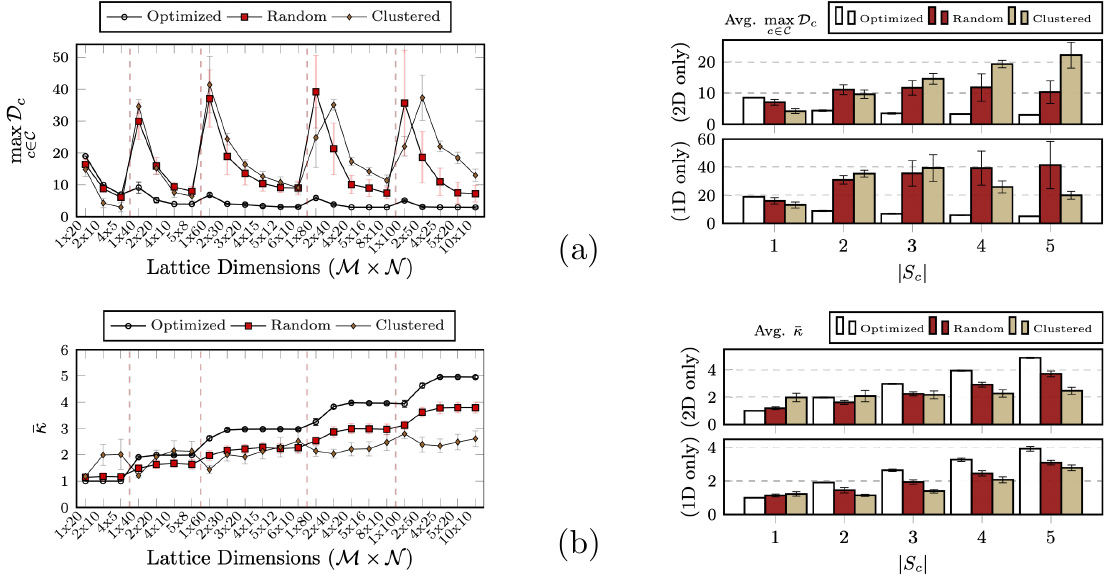}
    \caption{\justifying Static analysis of different 1D and 2D entanglement topologies with $20$ nodes. The plots show (a) the average worst-case inter-node distances ($\max_{c \in \mathcal{C}} \mathcal{D}_c$) and (b) the average number of vertex-disjoint inter-node paths ($\bar \kappa$), both computed over 100 independent executions for each lattice configuration and different qubit allocation strategy. The number of allowed qubits per node $\square_c = \mu |S_c|$ is constant within each group of lattices delimited by the dashed red vertical lines, while the histograms average the results for every allowed 1D and 2D lattice with increasing number of occurrences $|S_c|$.}
    \label{fig:05}
    \hrulefill
\end{figure*}

We assume that the decorations are performed by considering two additional qubits at each side (say left and right for the sake of exemplification) of each layer of $\mu$ parallel qubits in the resource state associated graph. 

We recall that the connectivity graph $\mathcal{\tilde G}$ remains the same with or without decorations, given the parallelism factor $\mu$. Hence, the explicit use of qubit decorations adjusts the number of physically required qubits to map the connectivity graph to the resource state graph, from $\mu$ to $3\mu$. In other words, as represented in Fig.~\ref{fig:01}a and Fig.~\ref{fig:04}a, every logical system $\tilde v \in \mathcal{\tilde V}$ is associated with a block of $\mu\times 3$ qubits of the shared resource state.

As a result, for a general resource state and given the entanglement topology $\mathcal{T}=(\mathcal{\tilde G}, f)$, the total number of qubits stored at each node is given by three layers of qubits stored in parallel, thus the total number after decoration becomes: $\hat \square_{c_i} = 3\square_{c_i}=3\mu |S_c|$. Other decoration strategies are possible, adding more qubits also in the vertical direction (decoration block), resulting in other expressions of $\hat \square_{c_i}$.

Thus, we can define the \textit{decorated} 2D entanglement topology $\mathcal{\hat T_{M,N}}$ as follows:
\begin{align}
\mathcal{\hat T_{M,N}} = 
    \bigl( 
    \hat M, 
    \hat N,  
    3\mu, 
    \square_c, 
    f 
    \bigr),
\end{align}
where the symbols $\hat M$ and $\hat N$ denote the parameters of the corresponding non-decorated topology $\mathcal{T_{M,N}}$ after the decoration, while the factor $\square_c$ is calculated by taking the maximum value over every $\hat \square_{c_i}$. 
Similarly, the decorated snake entanglement topology is generally defined as
$\mathcal{{\hat S}_{N}} = \bigl( \hat N, \, 3\mu, \square_c, \, f \bigr)$. 

A summary of the introduced notation and key design parameters is presented in Table~\ref{tab:key_parameters}.

\begin{figure*}[t]
    \centering
    \includegraphics[page=1,width=0.8\textwidth]{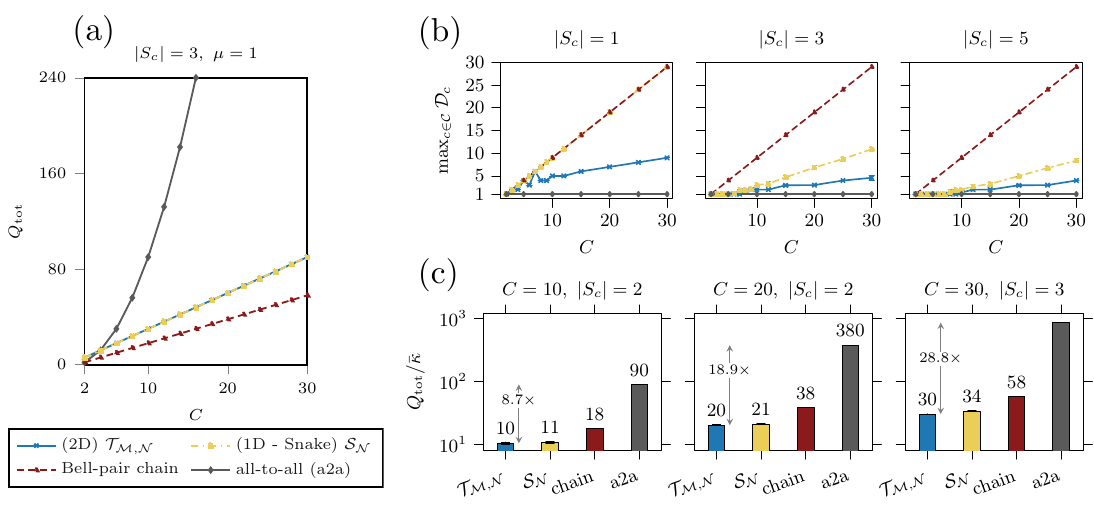}
    \caption{\justifying Comparison of the Snake $\mathcal{S_N}$ and 2D $\mathcal{T_{M,N}}$ allocations of our framework against the Bell-pair chain and the all-to-all configuration, without qubit decorations and optimized with simulated annealing over 5000 iterations. The plots show (a) the total number of required qubits ($Q_{\mathrm{tot}}$) vs. the number of nodes $C$, with $|S_c| = 3$ and $\mu = 1$, growing quadratically for the all-to-all configuration and linearly for the remaining schemes; (b) the worst-case inter-node distance ($\max_{c \in \mathcal{C}} \mathcal{D}_c$) vs. $C$, for $|S_c| = 1, 3, 5$, computed on $\mathcal{\tilde G}$ and hence independent of $\mu$; and (c) the qubit cost per independent path ($Q_{\mathrm{tot}}/\bar \kappa$) for three configurations of increasing size.}
    \label{fig:06}
    \hrulefill
\end{figure*}

\subsection{Static performance evaluation}

In the following, we evaluate the performance of the optimized qubit allocation strategy for entangled topologies in static and dynamic scenarios. The static analysis focuses on the first utilization of the resource state, showcasing not only the optimization of the inter-node distances but also an increase of the number of vertex-disjoint inter-node paths, directly related to the possibility of extracting dedicated communication links. Conversely, the dynamic analysis showcases the resilience of the $\mathcal{T_{M,N}}$ topology in case of a sequence of nested node failures.

We consider the logical lattice graph $\mathcal{\tilde G}$ with dimensions $\mathcal{M \times N}$ as the colored graph subject to the function $f$, where each color corresponds to the assignment of a logical vertex to a network node. Hence, the number of colors equals the number of network nodes $|\mathcal{C}| = C$. Without loss of generality, the optimization focuses on the allocation of logical systems in the entanglement topology, ensuring that the problem formulation remains independent of the specific parallelism factor $\mu$ or the presence of decorations in the resource state.

A \textit{static} performance evaluation analyzes the performance of the optimized allocation in a static scenario, where the number of nodes is fixed and the maximum number of qubits per node $\square_c$ is considered as a network constraint. In other words, we consider a static network scenario in which the resource state has been distributed with a fixed qubit allocation strategy but not yet utilized for the fulfillment of network requests.

We consider different 1D or 2D entangled lattice-shaped topologies with a fixed number of nodes $C$ and a maximum number of qubits per node $\square_c$. Our static performance analysis includes the following two case studies:

We compare the optimized allocation with a completely random allocation and with a clustered allocation, e.g. by forcing bigger areas of the cluster state to be assigned to the same node.
We present the results of the performance evaluation in terms of the (optimized) worst-case inter-node distance $\max_{c \in \mathcal{C}}\mathcal{D}_c$ and of inter-node vertex-disjoint paths $\bar \kappa$. The results are shown in Fig.~\ref{fig:05}. In this comparison the per-node budget is enforced strictly on the optimized allocation only, the resulting comparison is thus conservative.
We observe that the optimized allocation provides lower worst-case inter-node distances compared to other standard allocations, with a descending trend with the increase of the maximum number of qubits per node $\square_c$.
Such a static analysis is decoration-independent as it is performed on the connectivity graph $\mathcal{\tilde G}$.
The number of inter-node vertex-disjoint paths also increases as a by-product of the min-max objective, indicating that the optimized allocation provides a better connectivity between nodes and an overall flexibility in the choice of the communication patterns.

Remarkably, one can observe how randomly allocating qubits already yields strong performance, outperforming the naive clustered assignment and approaching the results of the optimized strategy. This suggests that, in certain scenarios, a simple random allocation, -- avoiding the overhead of solving complex optimization problems -- can serve as a highly effective and practical alternative.

We compare the use of a 1D or 2D cluster state as network resource versus the use of Bell state only, distributed between each pair of nodes, i.e., \textit{all-to-all} connectivity scenario.
Within the bipartite entanglement paradigm, the \textit{all-to-all} configuration is at the same time the most favorable and the most demanding one, since it attains the maximum connectivity achievable with bipartite resources, at a quadratic memory cost. Accordingly, we adopt it here not as a competing scheme, but as the upper reference against which the multipartite resource is assessed.
To complete the comparison, we further include a linear \textit{Bell-pair repeater chain} as the standard linear reference, providing exactly $\mu$ vertex-disjoint inter-node paths independently of $|S_c|$ -- similar to all non-colored baselines. We likewise include the 1D Snake topology, which for parallelism $\mu$ coincides with $\mu$ parallel merging-based quantum repeaters \cite{RuiRamWal-25} and reduces exactly to the Bell-pair chain at $|S_c|=1$. The comparison thus spans, across the memory, inter-node distance, and connectivity-cost axes, a linear repeater chain, the Snake $\mathcal{S_N}$ and 2D $\mathcal{T_{M,N}}$ allocations of our framework, and the all-to-all bipartite configuration.
This analysis does not rely on the connectivity graph but takes into account the exact number of qubits stored at each network node, including qubit decorations.
Our evaluation, represented in Fig.~\ref{fig:06}(a), measures the number of required qubits $Q_{\mathrm{tot}} = \sum_i \square_{c_i}$ for the distribution of a single Bell state for each pair of nodes in a network, versus the storage needed for the allocation of an optimized 2D cluster state. Clearly, by distributing a Bell state for every pair, an \textit{all-to-all} connectivity is enabled, but the number of required memories for the allocation of the qubits grows quadratically with the number of nodes of the network.
We highlight in Fig.~\ref{fig:06}(b) that the 2D cluster $\mathcal{T_{M,N}}$ exhibits a shorter worst-case inter-node distance than the 1D Snake $\mathcal{S_N}$ as the network grows. At $|S_c|=1$ the Snake collapses onto its non-multipartite counterpart, the Bell-pair chain, while the 2D cluster already exploits its lattice geometry to remain substantially shorter. Both the chain and the all-to-all configuration are structurally flat in $|S_c|$, since they lack an optimizable per-node qubit budget. Their connectivity is set purely by $\mu$ and $C$, with $\bar\kappa=\mu$ and worst-case distances $C-1$ and $1$, respectively. The framework, instead, converts extra per-node qubits into additional independent paths, with $\bar\kappa$ saturating near the $\mu|S_c|$ structural bound.
Fig.~\ref{fig:06}(c) quantifies this asymmetry in terms of the qubit cost per independent path. Because the 2D cluster converts each additional per-node qubit into an additional path, such a cost is almost flat in the per-node budget and grows only linearly with the network size. For the Bell-pair chain and the all-to-all configuration, whose extra qubits serve additional peers rather than additional paths to the same pair, the cost is governed by $C$ alone, growing linearly and quadratically, respectively. We stress that the panel does not aim at showing a connectivity advantage of the proposed resource over the all-to-all configuration, the latter attaining the minimum achievable distance at arbitrary parallelism. Rather, it makes explicit the price of such a minimum, which widens with the network size from $8.7\times$ to $28.8\times$, and which, rather than any connectivity deficit, is what renders the all-to-all configuration unviable at scale.

\subsection{Resilience analysis}

\begin{figure*}[t]
    \centering
    \includegraphics[page=1,width=1\textwidth]{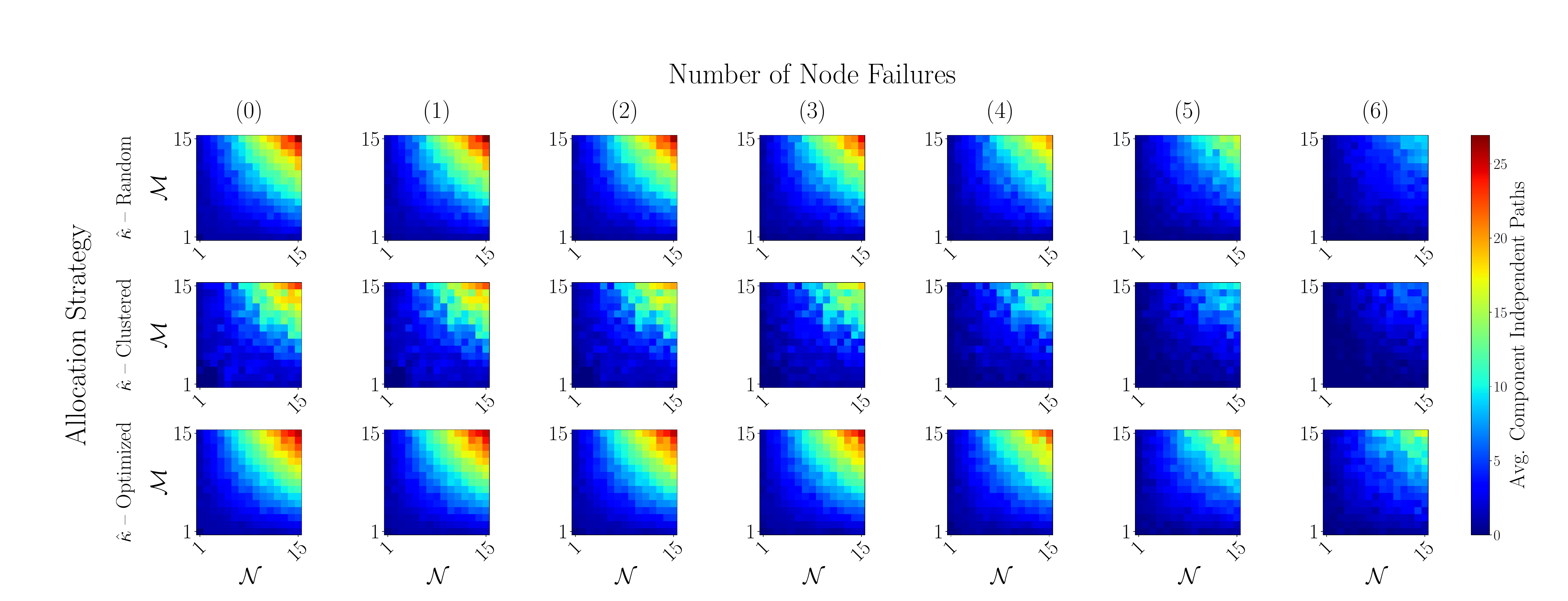}
    \caption{\justifying Impact of node failures on the average number of inter-node vertex-disjoint paths (calculated with respect to each graph component $\hat{\kappa}$) under different qubit allocation strategies, with the number of failed nodes increasing from $0$ (initial failure-free network) to $6$. The analysis is performed on $\mathcal{{T}_{M,N}}$ entangled topologies involving 8 nodes, with lattice dimensions $\mathcal{M} \times \mathcal{N}$ ranging from $1\times1$ to $15\times15$. Intra-node links are active and results are averaged over 25 independent realizations.}
    \label{fig:07}
    \hrulefill
\end{figure*}

In the following resilience analysis, we examine a sequence of nested node failures across various $\mathcal{T_{M,N}}$ topologies. Intrinsically, cluster states are resilient to the accidental measurement of part of the state, by measuring out every neighbor of each faulty qubit. 
In the following, we analyze the entanglement topology observed after a sequence of failures. 
More in detail, we account for both the formation of disjoint graph components and the potential activation of additional links between qubits stored at the same node.
Interestingly, an optimized qubit allocation strategy showcases solid performances in terms of worst-case inter-node distance $\max_{c \in \mathcal{C}} \mathcal{D}_c$ and average number of vertex-disjoint inter-node paths $\bar \kappa$. We recall that, as depicted in Figs.~\ref{fig:04}a and~\ref{fig:04}b, in case of a sequence of network failures the resulting entangled graph $\mathcal{ \tilde G}$ may be decomposed in disjoint connected components $K_i$. 
In order to formally consider the contribution of each connected component and its size, in our evaluation each connected component contributes to its own worst-case distance and average number of vertex-disjoint paths.

\begin{remark}
    According to Def.~\ref{def:connected_component}, if the graph is composed by different connected components $K_i \subseteq G$, for instance, after the failures of a subset of network nodes, then the number of inter-node vertex-disjoint paths can be calculated accordingly.  
    Let $\mathcal{C}_{K_i}$ be the set of colors present in $K_i$, then the average number of inter-node vertex-disjoint paths in \( K_i \) is given by:
    \begin{equation}
        \bar \kappa_{K_i} = \frac{2}{|\mathcal{C}_{K_i}|(|\mathcal{C}_{K_i}| - 1)} \sum_{c, c' \in \mathcal{C}_{K_i}: c < c'} \kappa(c, c').
    \end{equation}
\end{remark}

\begin{figure*}[t]
    \centering

    \includegraphics[page=1,width=1\textwidth]{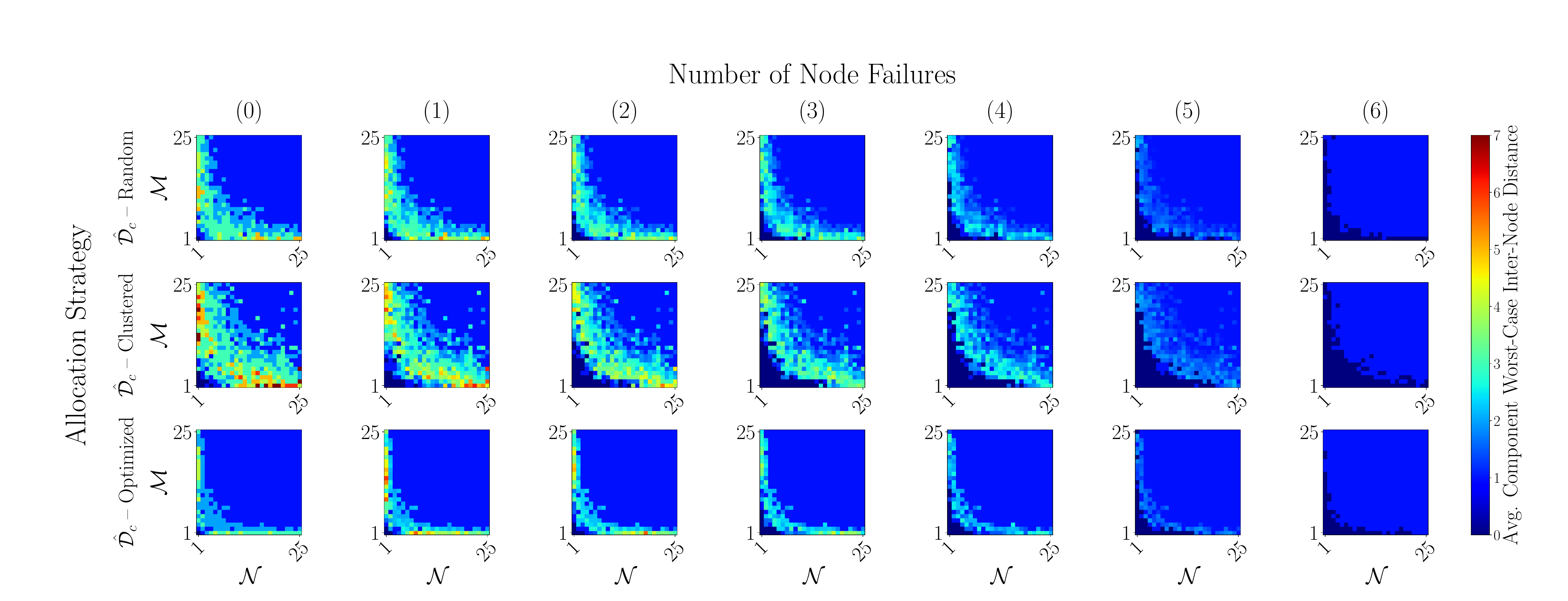}
    \caption{\justifying Impact of node failures on the worst-case inter-node distance (calculated with respect to each graph component $\mathcal{\hat D}_c$) under different qubit allocation strategies, with the number of failed nodes increasing from $0$ (initial failure-free network) to $6$. The analysis is performed on $\mathcal{{T}_{M,N}}$ entangled topologies involving 8 nodes, with lattice dimensions $\mathcal{M} \times \mathcal{N}$ ranging from $1\times1$ to $25\times25$.
    The worst-case inter-node distance is undefined for graph components containing a single network node: the configurations in which no multi-node component survives are reported as $0$ as a plotting convention.    }
    \label{fig:08}
    \hrulefill
\end{figure*}

Hence, a generalized derivation of the average number of inter-node vertex-disjoint paths also takes into account the set $K$ of connected components: $\hat \kappa =  \tfrac{1}{|K|}\sum_{i \in \{1,\dots,|K|\}} \bar \kappa_{K_i}$. 

Similarly, the average component worst-case inter-node distance can be calculated given the graph components $K_i$: $\mathcal{\hat D}_c = \tfrac{1}{|K|}\sum_{i \in \{1,\dots,|K|\}} \max_{c \in \mathcal{C}_{K_i}} \mathcal{D}_c$.

The resilience analysis reported in Fig.~\ref{fig:07} and Fig.~\ref{fig:08}, represents a faulty network scenario with 8 nodes using qubit decorations, and the values of $\mathcal{\hat D}_c$ and $\hat \kappa$ for every lattice configuration ranging from $1\times1$ (including the corresponding one dimensional entangled topologies) to $15\times15$, and $1\times1$ to $25\times25$. We observe the trend and the robustness of the average statistics -- referred to the disjoint connected components $K_i$ -- with an increasing number of node failures on the same entanglement topology, highlighting the differences between the optimized qubit allocation strategy introduced in \textit{Qubit assignment problem} and the allocation strategies discussed in \textit{Static performance evaluation}, namely, clustered and random.
We can notice that the optimized allocation strategy outperforms the clustered strategy in both worst-case distance and number of vertex-disjoint paths, for every number of failed nodes. Interestingly, the optimized allocation strategy provides overall better results than a fully random allocation, that is still a solid solution to limit network failures. 

Overall, we can note that the worst-case distance for the optimized strategy is more resilient than that of the fully random allocation when it comes to low error regimes (fewer than 50\% of the nodes fail). Conversely, optimized allocation guarantees a solid number of vertex-disjoint paths also with a larger number of node failures, showcasing overall better performance than clustered and random allocation across the explored failure range.
In general, the optimized allocation strategy enhances the error resilience of the entanglement topology, also guaranteeing useful entanglement when the number of failed nodes increases. 
Surprisingly, also a randomized allocation is sufficiently resilient to node failures, making the random allocation of the qubits a valid alternative to an optimization process every time the physical topology of the network updates, and the optimization problem cannot be solved in time for an incoming set of network requests. 
A typical use case for the randomized allocation could be the joining/deletion of a network node forcing a new optimization problem to be solved in parallel to the arrival of new requests to be fulfilled.

\section{Discussion}

In this work, we presented a generalized framework for entangled network topologies, enabling the description of the flexible allocation of the qubits of any graph state to the network nodes. 
Thanks to a proper engineering of the entangled network topology, the limitations induced by the physical network topology can be overcome, and the unconventional properties of multipartite entangled states can be exploited, with the nontrivial qubit allocation leading to important connectivity and robustness benefits.

Specifically, we provided a general framework for the optimized utilization of cluster states as network resources. Within this framework, we modeled our resource state defining a lattice-shaped entanglement topology $\mathcal{T_{M,N}}$, or, the simpler snake topology $\mathcal{S_{N}}$. According to the network storage constraints, this work showcased how a wise allocation of the qubits of the resource state to the network nodes corresponds to a reduction of the inter-node distances as well as enhanced failure resilience. Interestingly, the 2D cluster state is intrinsically a failure resilient resource, by taking into account qubit decorations. Indeed, a completely random qubit allocation guarantees fair inter-node distances and vertex-disjoint paths if the optimization problem cannot be solved, for instance while a new pool of requests arrives and the optimization problem is not yet solved. 

The unconventional properties of such multipartite entangled states find potential applications in quantum core networks, thus constituting a promising entangled backbone for quantum communication networks of the future.
Core network nodes must be capable of generating and distributing entangled resources according to optimized qubit assignment rules, allowing for a failure resilient agglomeration and routing of network requests through the generation of end-to-end entangled links across different clustered networks. 
Thanks to the proactive generation strategy, the joining or permanent disconnection of a core network node can be addressed at the next round of generation of the resource state whose allocation problem would take into account a different number of network nodes. Given the need to solve a new allocation problem, random allocation is still a beneficial alternative to fulfill incoming requests and minimize delays during optimization time. Both the random allocation and the greedy initialization of the heuristic (\textit{Numerical methodology} in Methods) can thus cover the optimization interval, serving incoming requests while a new allocation is computed. Neither of them matches the optimized allocation, but both keep the network operational instead of interrupting the service.

Given the generality of the proposed approach, any other allocation strategy can be used, as well as an arbitrary built resource state. We plan to further explore the utilization of different resource states, highlighting their potential in the context of quantum communication network at different scales. 
Remarkably, different resource states can be built by taking into account an a priori optimal allocation, enforcing logarithmic scaling of the distances between the nodes and thus building a tree-like resource state. Similarly, the average distances between remote nodes can also be shortened by considering three-dimensional (3D) or even $k$-dimensional ($k$D) cluster states, also enforcing the concept of vertex-disjoint paths as indicators of number of extractable Bell states.
Generally speaking, an optimized resource state can be tailored to the specific problem to be solved, prioritizing flexibility of the manipulations (cluster states) or minimizing the distances or even giving much more importance to distribution simplicity, in the context of short- and mid-time-horizon implementations.

We finally observe that our framework, for all its generality, operates at a \textit{structural} rather than at the physical resource level: it abstracts the imperfections involved in handling the multipartite resource, and models failures at the granularity of qubits becoming unavailable at the network nodes. Taking the framework proposed here as the starting point, its natural continuation is the exploration of further forms of optimization and of additional exploitable structural properties, alongside the resource states discussed above.

We hope that this work will promote the search for increasingly realistic and optimized multipartite entangled resource states and fuel the interest of the community for the design of future quantum networks.

\section{Methods}

\subsection{Numerical methodology}

\begin{figure}[t]
    \centering
    \includegraphics[page=1,width=0.36\textwidth]{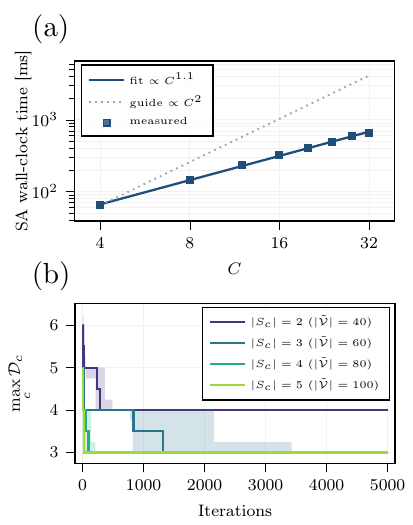}
    \caption{\justifying Runtime (a) and convergence (b) of the allocation heuristic ($T_0 = 10$, cooling rate $0.99$, greedy balanced initialization). The runtime fit gives $\propto C^{1.1}$, against a $\propto C^2$ guide anchored at the smallest $C$. Panel (b) spans the full adopted $5000$-iteration budget. No run improves beyond iteration $\approx\!3500$, and the flat tail confirms that the adopted budget is a safe margin.}
    \label{fig:sa_conv_runtime}
    \hrulefill
\end{figure}

All the results reported in this work are obtained with a custom graph-theoretic simulation. Inter-node distances (Def.~\ref{def:inter_color_distance}) are computed by multi-source breadth-first search, while the number of vertex-disjoint inter-node paths (Def.~\ref{def:vertex_disj_inter_color_paths}) is obtained exactly, as a maximum flow on the node-split graph.
The optimized allocation is obtained with the simulated annealing (SA) introduced in \textit{Qubit assignment problem} ($T_0 = 10$, cooling rate $0.99$, $5000$ iterations), initialized with a balanced greedy assignment that respects the per-node qubit budget $\square_c$. At each iteration, the algorithm evaluates the objective, namely the worst-case inter-node distance, over the logical connectivity graph $\mathcal{\tilde G}$, whose vertices are the $|\mathcal{\tilde V}|$ \textit{logical systems} and whose $C$ colors label the network nodes. A swap move exchanges the colors of two logical systems and hence changes only the two corresponding rows and columns of the $C\times C$ matrix of inter-color distances $d(c,c')$, leaving the other $C-2$ colors untouched. The objective is thus updated \emph{incrementally} by refreshing the two affected rows with two multi-source shortest-path searches, $O(|\mathcal{\tilde V}|)$ each, followed by an $O(C^2)$ scan of the matrix.
As a consequence, the heuristic runs at a per-iteration cost $O(|\mathcal{\tilde V}| + C^2)$.  Thus, over $I$ iterations, and given the relation between nodes and their occurrences ($|\mathcal{\tilde V}| = C\,|S_c|$), the total cost is $O\big(I\,(C\,|S_c| + C^2)\big)$, while the total memory cost is $O(|\mathcal{\tilde V}| + C^2)$. Crucially, the optimization runs on the logical graph $\mathcal{\tilde G}$ rather than qubit by qubit, so its cost is independent of the parallelism factor $\mu$ and of the presence of decorations: enlarging each logical system with additional physical qubits leaves the allocation runtime unaffected, since batches of $\mu$ qubits are mapped to the same logical system and hence to the same color in the assignment process.

Runtime and convergence are reported in Fig.~\ref{fig:sa_conv_runtime}, as single-threaded wall-clock times on an Apple M2 (8 cores, 16~GB). Interestingly, for the reference networking use-case, i.e., a network core of a few tens of nodes, the two shortest-path searches dominate the $O(C^2)$ matrix scan in wall-clock terms, by more than an order of magnitude over the range of interest, the two contributions crossing over only far outside it. Hence, the measured wall-clock time grows almost linearly with $C$: a power-law fit over $C\le 32$ yields an exponent of $1.1$ (Fig.~\ref{fig:sa_conv_runtime}(a)), the residual quadratic term becoming appreciable only well beyond this range. The objective, in turn, plateaus well within the adopted budget (Fig.~\ref{fig:sa_conv_runtime}(b)), the median plateau being reached within a few hundred iterations, and below $\approx\!800$ in every configuration considered. The size of the budget is nonetheless required to accommodate the dispersion across runs, individual seeds still improving as late as iteration $\approx\!3500$.

\section*{Acknowledgements}
F. Mazza and J. Illiano acknowledge PNRR MUR NQSTI-PE00000023.
A. S. Cacciapuoti's and M. Caleffi's work has been funded by the European Union under the ERC grant QNattyNet, n.101169850. Views and opinions expressed are however those of the author(s) only and do not necessarily reflect those of the European Union or the European Research Council. Neither the European Union nor the granting authority can be held responsible for them.
This work was also funded in part by the Austrian Science Fund (FWF) 10.55776/P36009, 10.55776/P36010. For open access purposes, the author has applied a CC BY public copyright license to any author accepted manuscript version arising from this submission. Finanziert von der Europäischen Union NextGenerationEU.

\section*{References}
\bibliographystyle{naturemag}
\bibliography{biblio}

\end{document}

%% file: preamble.tex
\usepackage[utf8]{inputenc}
\usepackage[T1]{fontenc}
\usepackage[english]{babel}
\usepackage{amsmath, amssymb}
\usepackage[bbgreekl]{mathbbol} 
\usepackage{braket}          
\usepackage{graphicx}
\usepackage{caption} 
\usepackage{array}[=2016-10-06]
\usepackage{ragged2e}
\usepackage{amsthm}
\usepackage[dvipsnames]{xcolor}

\usepackage{hyperref}

\hypersetup{
    colorlinks = true,
    linkcolor = BlueViolet,
    citecolor = BlueViolet,
    filecolor = BlueViolet,
    urlcolor = BlueViolet,
}

\usepackage{geometry}
\theoremstyle{plain}

\theoremstyle{definition}
\newtheorem{definition}{Definition}

\theoremstyle{remark}
\newtheorem*{remark}{Remark}